# Towards a Trusted Execution Environment via Reconfigurable FPGA

Sérgio Pereira
Universidade do Minho
sergio.pereira@dei.uminho.pt

David Cerdeira
Universidade do Minho
david.cerdeira@dei.uminho.pt

Cristiano Rodrigues
Universidade do Minho
id9492@alunos.uminho.pt

Sandro Pinto
Universidade do Minho
sandro.pinto@dei.uminho.pt

*Abstract*—Trusted Execution Environments (TEEs) are used to protect sensitive data and run secure execution for security-critical applications, by providing an environment isolated from the rest of the system. However, over the last few years, TEEs have been proven weak, as either TEEs built upon security-oriented hardware extensions (e.g., Arm TrustZone) or resorting to dedicated secure elements were exploited multiple times. In this project, we introduce Trusted Execution Environments On-Demand (TEEOD), a novel TEE design that leverages the programmable logic (PL) in the heterogeneous system on chips (SoC) as the secure execution environment. Unlike other TEE designs, TEEOD can provide high-bandwidth connections and physical on-chip isolation. We implemented a proof-of-concept (PoC) implementation targeting an Ultra96-V2 platform. The conducted evaluation demonstrated TEEOD can host up to 6 simultaneous enclaves with a resource usage per enclave of 7.0%, 3.8%, and 15.3% of the total LUTs, FFs, and BRAMS, respectively. To demonstrate the practicability of TEEOD in real-world applications, we successfully run a legacy open-source Bitcoin wallet.

## I. INTRODUCTION

Security is becoming paramount for IoT end-to-end solution designs [1]. One well-established strategy to provide increasing integrity and confidentiality for applications, from the edge to the cloud, relies on Trusted Execution Environments (TEE). TEEs drastically reduce the trusted computing base (TCB) of the systems by providing a secure execution environment for security-critical applications that are isolated from the operating system or the hypervisor [2]–[4]. Applications loaded onto the TEE are guaranteed to run and process data in a secure environment, also known as enclaves, isolated from the rest of the host system, i.e., the rich execution environment (REE). We depict this definition in Figure 1. Private user data is stored in a secure storage area, and sensitive functions are executed inside a TEE without interference from the REE. Therefore, even if attackers have full control over the REE, in principle, they cannot corrupt or leak data processed and stored inside a TEE.

One of the most common TEE design approaches is to create a virtual secure processor in the main application processor by leveraging specific security-oriented hardware extensions. Intel SGX [5] and Arm TrustZone [6] are prominent examples of such technologies, widely used in the context of cloud and mobile applications, respectively. However, both of these approaches yield different weaknesses. TrustZone and SGX have been successfully attacked multiple times, with highly damaging impacts across various platforms, casting doubts on the effectiveness of the security guarantees that existing commercial TEEs can, in practice, provide [7]–[15].

An alternative approach that has also been taken by industry to provide a TEE relies on dedicated external secure elements. Google's Titan [16] and Apple's T2 [17] implements a secure element by externally mounting a dedicated security processor next to the main CPU. While dedicated secure elements provide strong isolation between the REE and the TEE, the off-chip communication fabric is physically exposed to an attacker, making it vulnerable to probing attacks [18]. Furthermore, the slow communication interface between the two domains, e.g., SPI communication for Google Titan, limits the applicability to use cases with high-bandwidth requirements (e.g., Digital Rights Management services).

### A. Contributions

In this project, we introduce a novel TEE design aiming at disrupting the way TEEs are being built and deployed. We propose a newly refined TEE approach, named Trusted Execution Environments On-Demand (TEEOD), which leverages reconfigurable FPGA technology to provide additional security guarantees for security-critical applications. TEEOD implements secure enclaves in the programmable logic (PL) by instantiating a customized and dedicated security processor per application on a per-need basis. The main reason why the PL is seen as a suitable enabler for this purpose is that it is physically isolated from the would-be malicious main CPU. Therefore, the PL can act as a TEE as long as proper control mechanisms are implemented to regulate arbitrary accesses. The TEEOD APIs are compliant with the GlobalPlatform API specifications to provide interoperability while deploying and managing legacy trusted applications (TAs). To demonstrate the practicability of TEEOD in real-world applications, we have deployed an open-source hardware Bitcoin wallet on our proof-of-concept implementation. We have also conducted our evaluation on a Xilinx Zynq UltraScale+ MPSoC development board (Ultra96-V2), focusing on hardware costs and performance. The assessed results demonstrate that TEEOD can run up to 6 enclaves simultaneously, with a resource usage per enclave of 7.0%, 3.8%, and 15.3% of the total LUTs, FFs, and BRAMS. TEEOD outperforms the requirement of 2 simultaneous active TAs from the standard GlobalPlatform specification.

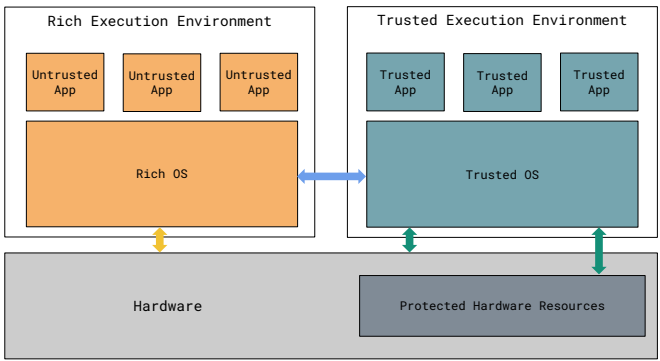

Fig. 1. Overview of a TEE example. On the right side the TEE (blue), the trusted area where trusted applications execute. On the left side the REE (orange), the untrusted area where typically a rich OS and untrusted applications execute.

## II. THREAD MODEL AND REQUIREMENTS

### A. Threat Model

Our threat model is mainly inspired by the TEE attack scenarios defined by Cerdeira et al. [19], i.e., we consider that an attacker can take control of the REE and the interfaces to the TEE components. Additionally, we consider that the secure world software in TrustZone is malicious. This way an attacker has full access to the platform's buses, taking into account that the hard CPU available in the PS can access them. Of particular importance is that an attacker may try to reconfigure the FPGA hardware; however, it's possible to guarantee that the bitstream installed in the FPGA is authentic and cannot be replaced, by continuously monitoring for any backdoor activity (e.g., JTAG and FPGA configuration ports) [20], [21]. In addition, we consider that TEEOD TAs can be malicious, and once deployed in the system, they may try to steal security-sensitive data from other TAs or otherwise attack the host platform by attacking secure devices or accessing protected memory regions.

### B. TEEOD Requirements

A TEE must provide a set of essential security features [22]. These features encompass the main building blocks to build secure execution environments for performing security-critical operations and data processing.

*Isolation from the REE.* TEEOD must provide isolation from the REE executing in the PS. This requirement prevents malicious untrusted applications from accessing and compromising TAs (executing in TEEOD).

*Isolation from other TAs.* TEEOD must provide isolation between multiple TAs. This requirement prevents malicious or buggy TAs from accessing and compromising other TAs.

*Application management control.* TEEOD must only instantiate and allow modifications of TAs from pre-approved trusted sources. This requirement provides confidence in the trustworthiness of TAs.

*Identification and binding.* TEEOD must leverage SoC built-in mechanisms for binding data to the platform, thus preventing malicious agents from accessing data outside of the platform.

*Trusted storage.* TEEOD must guarantee the integrity, confidentiality, and device binding of data belonging to TAs and the TEE. This requirement enables TAs to securely store secrets or other security-critical data in a secure memory area.

*Trusted access to peripherals.* TEEOD must provide the ability to securely pair devices and TAs, thus allowing for trustworthy interaction between them.

*State of the art cryptography.* TEEOD must feature cryptographic algorithms (e.g., AES, RSA, ECC) and commonplace cryptographic primitives such as random number generation (RNG) and monotonic time stamps. This requirement guarantees that the mechanisms built atop TEEOD are resilient to known cryptographic attacks and weaknesses.

## III. DESIGN

### A. Design Principles

In light of the aforementioned threat model and requirements, we relied on a set of principles to guide our design. These design principles intend to avoid widely-known security pitfalls that have affected TEE systems, i.e., TrustZone-assisted TEEs, over the years.

We followed a set of architectural principles aiming at distributing responsibilities of system components, their level of privilege concerning other components, and how and when these components interact with other parts of the system. As found in the literature [10], [15], [23]–[29], shared hardware resources have repeatedly been leveraged to extract the secrets and manipulate core TA functionality. Sharing the software runtime has also lead to the compromise of critical security functions [30]–[32]. In TEEOD, we strive to **avoid sharing hardware and software resources**. In addition, in the TEEOD architecture, we adopt the **principle of least privilege**, i.e., each component is given as little privilege as needed to perform its functions. For example, in TEEOD, TAs cannot arbitrarily access main memory. The same happens for the interaction between components, where each component has **well-defined and carefully designed interfaces** so that unexpected interactions between components do not compromise system security.

### B. Design Overview

As seen in Figure 2, in the overall TEEOD architecture, TEEs are materialized through secure enclaves instantiated on the reconfigurable hardware. These enclaves can host custom accelerators or soft generic processors depending on user requirements. For the sake of this project, we emphasize the design implications of soft-processors as enclaves, in an effort to foster portability and interoperability among legacy TAs. So, the use case for custom accelerators is currently out-of-scope. From the point of view of the REE application, the flow follows the default RPC-model, i.e., the client application issues the request and waits for the reply from the TA.

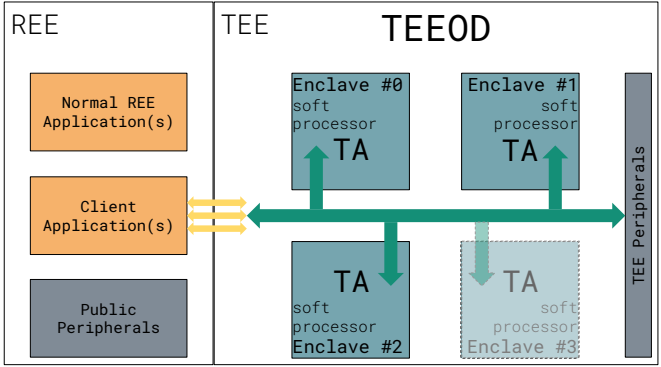

Fig. 2. Depiction of the TEEOD overall design. On the left side of the figure (orange) the untrusted applications. On the right side of the figure (blue) are the secure enclaves. At green and yellow are the arrows representing the secure communication channel between REE and TEE.

### C. Enclave Life-cycle

*Enclave Creation.* TEEOD keeps a list of the available enclaves (*enclaves_list*) and a list of the loaded TAs (*loaded_tas*). For an REE application to communicate with a particular TA, it needs first to call the TEE Client API to connect with the TEEOD. Along with the information that the REE application wants to communicate, the TEE Client API also keeps the TA's universally unique identifier (UUID) and their location on the continuous physical memory (CM). Then, TEEOD checks if the TA was already loaded to an available enclave. If the TA is not loaded, TEEOD copies the target TA binary to a private Tightly-Coupled Memory (TCM) available per enclave. Once the loading process is complete, TEEOD marks the enclave as taken and the TA as properly loaded. Afterward, the TA enters in wait for interrupt state (WFI) until a communication request is issued from the REE application.

*Enclave Execution.* When TEEOD completes the creation of the enclave, the TA keeps waiting for a signal to start executing. So, immediately upon the REE application sends a command, through the previously established session with TEEOD, the TA wakes up and starts executing the trusted code. The typical flow is as follow: REE sends the information that wants to communicate with the desired TA to the TEEOD; TEEOD signals the enclave through an interruption, notifying that a new message has arrived; TA reads the message from the shared registers and performs the command requested by the REE; the TA handles the command, and notifies the TEEOD that it has finished the requested command by clearing a shared register; finally, the TA goes back into the WFI state.

*Enclave Destruction.* Once the session between the REE application and the TA completes, i.e., calling the close session function provided by the TEE Client API, TEEOD notifies the TA that the session shall be closed. This notification is similar to sending any other message to the TA. After the TA processes the command, TEEOD clears the memory region of the enclave, hard resets the soft-core and cleans its microarchitectural states, marks the enclave as available on the *enclaves_list*, and removes the TA's UUID from the *loaded_tas*, ensuring no information is leaked while running a new TA in that enclave.

### D. Communication handling

REE applications can interface, i.e., exchange information via raw data or shared memory, with TAs by leveraging the TEE Client API. The REE communication agent (CA) module is responsible for issuing the requests and sending the respective information. The REE CA passes the TEEOD (i) the UUID of the target TA, (ii) the requested message in form of shared registers between the REE and TEEOD, (iii) the message type, and (iv) the access permissions that the TA has to the message. Then, the TEEOD validates the received information, and in case of success, sends the message to the desired TA through shared registers between TEEOD and the enclave, and triggers the enclave to wake up from WFI state. When the TA finishes the requested service, it always replies whether its operation was performed successfully or not. The reply is transmitted on the same shared registers used to receive the message.

### E. Shared Memory Management

Shared Memory is a block of memory that is shared between the TEE and the REE and it is typically used to transfer large blocks of data between them. Shared memory communication is handled by the REE CA, i.e., it is responsible for first allocating the requested memory. Then, provide the address and size of the shared memory area in use. The communication handling of the shared memory's pointer and size is made by the same mechanism as any other type of message.

## IV. PoC Implementation

In this section, we describe the PoC implementation of the TEEOD architecture. We first explain the implemented hardware blocks and then we overview the TEE Client API and the TEE Internal Core API implementation.

### A. TEE Hardware Blocks

The current PoC implementation takes advantage of heterogeneous multiprocessing (PS+PL) architectures. TEEOD is mostly implemented on the PL, and expects the REE to run in the Application processors of the PS. As illustrated in Figure 3, the main hardware blocks are TEE Manager Agent IP, TA Loader Agent IP, TEE Communication Agent IP, soft processors for the enclaves, and BRAMs for TCM of the soft processors.

*TEE Manager Agent IP.* The TEE Manager Agent IP (TEE MA) is responsible for controlling the execution of TA Loader Agent IP and TEE Communication Agent IP. These modules are responsible for managing the creation and destruction of enclaves, as well as the correct execution of each TA. TEE MA has two lists, *enclaves_list* and *loaded_tas*. The *enclaves_list* keeps track of the available enclaves, where each enclave is identified by the address of the respective private TCM. The *loaded_tas* list saves the UUID of each running TA and the respective enclave. Implemented as an AXI4 Slave IP, the TEE MA shares with the REE the following registers: address, size, and UUID of the intended TA; status register to indicate the state of loading of the TA. When TEE MA receives these registers, it first checks if the UUID is already listed on the *loaded_tas*. If the UUID is not listed, it picks one of the available enclaves and sends the address (*addr_src* output register) and size (*size* output register) of the intended TA to the TEEOD Loader Agent IP, as well as the TCM address of the chosen enclave (*addr_dst* output register). After the TEEOD Loader Agent IP complete the transfer of the TA binary to the chosen enclave, the TEE MA marks the enclave as taken and communicates to the REE, via the status register, that acknowledges that the loading operation was completed. The TEE MA also communicates, via a wired signal, to the TEE Communication Agent IP that the TA was currently loaded and can start to communicate. Each enclave has a reset signal that is connected to the TEE MA, so the TEE MA has full control over the enclaves.

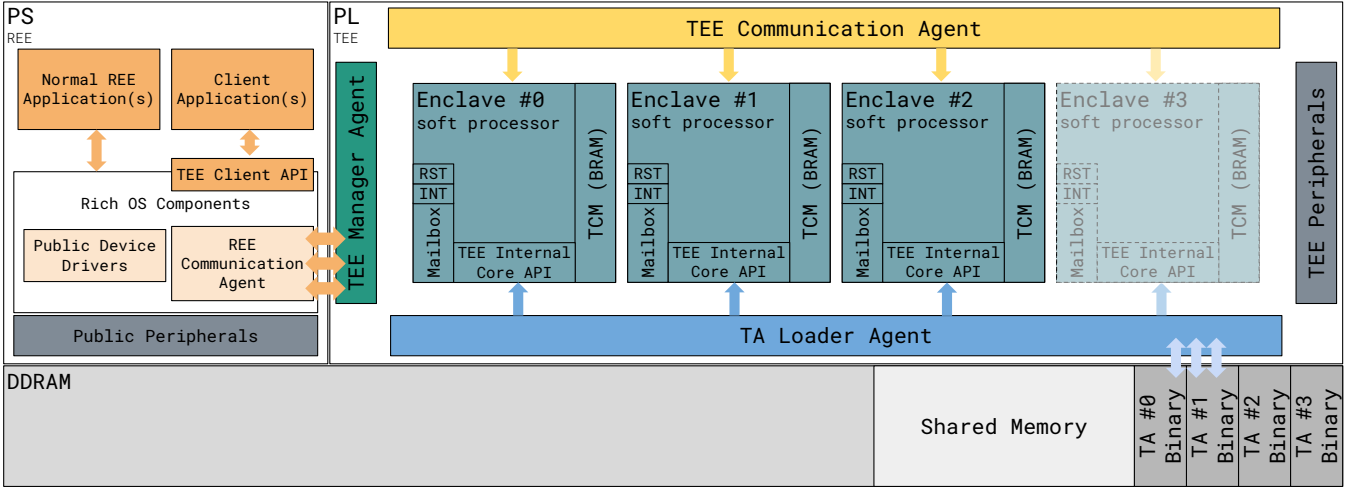

Fig. 3. Representation of the TEEOD Architecture. On the left side (orange), the untrusted applications and the rich components. On the right side (turquoise), four secure enclaves. Of which three of them are occupied and the other one, "Enclave #3", is available to be used. At the bottom (grey), the shared memory and the CM area where the TAs binaries are placed.

*TA Loader Agent IP.* The TA Loader Agent IP is responsible for loading TAs to the private memory of the enclave. The TAs binaries are stored in a large block of physical-contiguous memory, which is created at boot time by the rich OS (Linux) using the contiguous memory allocator (CMA). When a specific TA is needed, a region of memory within this physical-contiguous area is allocated to load the TA binary. Per TEE Manager Agent request, the TA Loader Agent receives the base address as well as the size of the region where the target TA encrypted binary is located. Then it triggers high-bandwidth directed memory access (DMA), loading the TA binary from the DDRAM to the target private enclave TCM, without any PS intervention. In the current TEEOD implementation, binaries are in plaintext and not signed. We did not implement the authentication, encryption, and trusted storage mechanisms. In the near future, we will extend the TEEOD to include all these security-related features.

*TEE Communication Agent IP.* The TEE Communication Agent (TEE COMM) is responsible for enabling communication between REE and TEE. It has a message box (Mailbox) for each active enclave. Enclaves do not access mailboxes from other enclaves, respecting the "*Isolation from other TAs*" requirement. TEE COMM has also an interrupt signal wired to each enclave to notify the enclave whenever there is a new available message. The mailbox is composed of these registers: *operation_id*, to identify if the operation is an open, close, or invoke; *session_id*, to identify the session that is calling; *param_type*, to identify the type of the 8 general-purpose registers that comes next; *gp_params*, general-purpose registers that depending on the *param_type* are shared memory address or raw data; *cmd_id*, general-purpose register to indicate the desired command, only needed for the invoke-command operation. Once the TEE COMM receive these registers from the REE application, it waits for the TEE MA reply, carrying information about the enclave in use. With this information, the TEE COMM starts to copy all the mailbox registers to the mailbox of the target enclave. After completing the copy, the TEE COMM sends an interrupt signal to the enclave. TEE COMM waits for the enclave to clear the interrupt, and copies the enclave's mailbox registers back to the REE's shared mailbox registers.

*Enclaves.* Enclaves are individual modules built around a dedicated lightweight soft processor (e.g., Cortex-M1), private BRAMs to store the TA code, mailboxes to receive and transmit information with REE, a shared memory space, and two interrupts: reset (RST signal) and TEE communication interrupt (INT signal). The interrupt reset is connected to the TEE Manager Agent IP and the TEE communication interrupt is connected to the TEE Communication Agent. When the RST signal is down, the enclave runs the TA present in the private BRAM. Whenever it receives the INT signal, the enclave leaves the WFI state and enters in the interrupt service routine (ISR). At the ISR, the enclave fetches all the registers available in the mailbox. Depending on the *operation_id* register, the enclave executes different operations. After executing the target operation, the enclave clears the interrupt and enters back into the WFI state.

### B. TEE Client API

The TEE Client API describes and defines how a client running in an REE should communicate with the TA running inside the TEE. To identify a TA to be used, the client provides a UUID. All TA's exposes one or several functions. Those functions correspond to a so-called *commandID* which is also sent by the client. From the GlobalPlatform TEE Client API Specification, we have implemented the following APIs: *TEEC_OpenSession*, *TEEC_InvokeCommand*, *TEEC_CloseSession*. The *TEEC_OpenSession* API opens a new Session between the CA and the specified TA and communicates to the TEE COMM the *operation_id*. When returning TEEC_SUCCESS, it means that TEE COMM has written an id to the *session_id* register. The *TEEC_InvokeCommand* API invokes a command within the specified Session. This API is also responsible for allocating a portion of the shared memory if the *param_type* argument points to an address of a shared buffer.

### C. TEE Internal Core API

The Internal Core API is the API that is exposed to TAs running in the secure area. The TEE Internal API consists of four major parts: (i) Trusted Storage API for Data and Keys; (ii) Cryptographic Operations API; (iii) Time API; and (iv) Arithmetical API. From the GlobalPlatform TEE Internal Core API Specification, we have only implemented the necessary features to successfully run the bitcoin wallet application i.e., some functions of the Trusted Storage API and some functions of the Cryptographic Operations API.

## V. USE CASE: BITCOIN WALLET

To demonstrate the practicality of TEEOD in real-world applications, we have tested an open-source bitcoin wallet application that was developed for OP-TEE. The Bitcoin wallet use case is interesting due to the increasing interest in cryptocurrency and blockchain technology in general, as well as for the use of security features that need to be supported. The wallet implementation uses a subset of features of the TEE internal core API, namely: random number generation for creating random seed phrases, secure storage to hold the master key, and cryptographic primitives to sign chain transactions and perform other cryptographic related operations.

### A. Bitcoin Wallet

The bitcoin wallet application consists of two parts, the client application (BW_CA), which executes in the REE (i.e., Linux) and the TA (BW_TA), which executes in TEEOD. The BW_CA makes use of the TEE Client API to access the following operations provided by BW_TA: checks if a master key already was generated (command_1); generates a new master key and corresponding mnemonic, i.e., seed phrase (command_2); generates a new master key from a given mnemonic (command_3); deletes a master key (command_4); signs a blockchain transaction (command_5); and obtains the bitcoin wallet address (command_6).

*Client Application* The BW_CA is a Linux application and uses the TEE Client API, thus, no modifications were necessary to run on our system. The BW_CA provides an interface to use the wallet and it is designed to perform one operation per invocation. On every invocation, the BW_CA opens a session with the BW_TA, invokes the command with the necessary arguments, and closes the previously opened session after the operation is completed. To successfully request one of the six functionalities implemented in BW_TA, the client application uses the following arguments: the command id of the desired functionality, a 4-digit access pin, and optional arguments depending on the functionality (e.g., a mnemonic to generate a master key).

*Trusted Application* The BW_TA is an application developed to run atop OP-TEE. However, because OP-TEE provides to its TAs a TEE Internal API library, the effort to execute the TA on TEEOD was narrowed to the compilation of the BW_TA with the TEEOD's TEE Internal Core API library. The library handles the requests and invokes the appropriate TA handlers (e.g., open session and invoke-command).

The BW_TA operates as follows. On opening and closing sessions, BW_TA simply returns success. On the invoke-command call, the BW_TA receives required parameters through the mailbox, including the command id, the 4-digit access pin, and additional command-specific arguments. Some arguments may need to be passed through shared memory, such as command_3 that needs to pass a seed phrase to generate a master key.

### B. Bitcoin wallet demonstration

Figure 4 shows the successful execution of the six possible wallet commands. Note that the enclave that hosts the BW_TA has a dedicated UART peripheral, which outputs the status of the enclave internal's operations. First, we check if a master key already was generated. Second, we generate a new master key from a given mnemonic phrase. Third, we sign a transaction. Fourth, we get the bitcoin receiving address. Lastly, we run the command to erase the generated key.

## VI. EVALUATION

To evaluate TEEOD, we have deployed our prototype on an Ultra96-V2, a development board based on the Linaro

Fig. 4. Bitcoin wallet demo running the six implemented functionalities: (i) on the left, an application running on top of the PetaLinux REE OS, which calls the Bitcoin wallet TA; (ii) On the right, the outputs on the enclave secure UART of the Bitcoin wallet TA replies.

96Boards Consumer Edition (CE) specification. The Ultra96-V2 features a Xilinx Zynq UltraScale+ MPSoC ZU3EG with 2 GB DDR4 SDRAM. All FPGA modules were implemented in Verilog-HDL, using Xilinx Vivado tool. Figure 5 depicts the block diagram of the developed system. The diagram shows the necessary blocks to implement a TEE with one enclave. The enclave features a UART peripheral. The soft processor embedded in the enclaves is an ARM Cortex-M1. This core is very similar to the Cortex-M0, highly optimized for FPGA implementation. Cortex-M1 is available as free to use via ARM DesignStart FPGA.

According to GlobalPlatform specification, a compliant TEE shall be able to host TAs with the minimum binary TA code of 64 kiB. For the sake of this evaluation, we configured the private TCM of each enclave to 64 kiB of RAM and 8 kiB of shared memory.

### A. Synthesis Results

We synthesized the TEEOD design onto the Zynq UltraScale+ ZU3EG with up to four enclaves. We evaluated the hardware logic required for each TEEOD design case by assessing the number of LUTs, flip-flops (FFs), digital signal processors (DSPs), and block RAMs (BRAMs). The synthesis results are shown in Table I. Comparing the resource utilization per design, we observe that each enclave adds to the system an average of 7.0% (5000/70560) of the total LUTs, 0.6% (180/28800) of the total LUTRAMs, 3.8% (5500/141120) of the total FFs, 15.3% (34/216) of the total BRAMs, and 0.9% (3/360) of the total DSPs. The most used resource for enclave synthesis is BRAM. This could become a bottleneck for the scalability of the TEEOD solution, limiting the number of active enclaves and TAs at a specific point in time. However, according to GlobalPlatform specification, a compliant TEE shall be able to host two TAs at the same time. So, in its current form and targeting the Ultra96-V2, TEEOD can run up to 6 enclaves without running out of BRAM resources. This is enough to comply with the standard GlobalPlatform specification. For future work, we will study how to securely attribute regions of main memory to enclaves, thus improving the scalability of the TEEOD solution.

TABLE I
SYNTHESIZED RESULTS OF FOUR TEEOD DESIGN CASES

| Resources | 1 Enclave | 2 Enclave | 3 Enclave | 4 Enclave |
|---|---|---|---|---|
| LUT | 9845 (13.95%) | 14844 (21.0%) | 20146 (28.55%) | 24963 (35.37%) |
| LUTRAM | 807 (2.80%) | 987 (3.4%) | 1171 (4.0%) | 1355 (4.7%) |
| FF | 11532 (8.17%) | 17034 (12%) | 22735 (16.11%) | 28026 (19.85%) |
| BRAM | 34 (15.7%) | 68 (31%) | 102 (47.22 %) | 136 (62.96%) |
| DSP | 3 (0.83 %) | 6 (1.6 %) | 9 (2.5%) | 12 (3.3%) |
| IO | 2 (2.4%) | 2 (2.4%) | 2 (2.43%) | 2 (2.4%) |
| BUFG | 3 (1.53%) | 4 (2.%) | 4 (2 %) | 4 (2%) |

### B. Performance Analysis

We next evaluated TEEOD in terms of performance. To obtain experimental results, each of the experiments was repeated 100 times. The power-saving mode was turned off and the CPU frequency was set to the maximum value to minimize variation between experiments.

First, we measured the execution time of the open session operation (*TEEC_OpenSession*) when a TA of 64 kiB is not loaded, which took about 54.4 milliseconds on average. Then, we measured the to open a session, of the same TA, when it is already loaded on one enclave. In this case, the execution time was 31.1 microseconds, 3 orders of magnitude shorter when compared to the previous case, because TEEOD keeps track of the available TAs and so does not need to load it again.

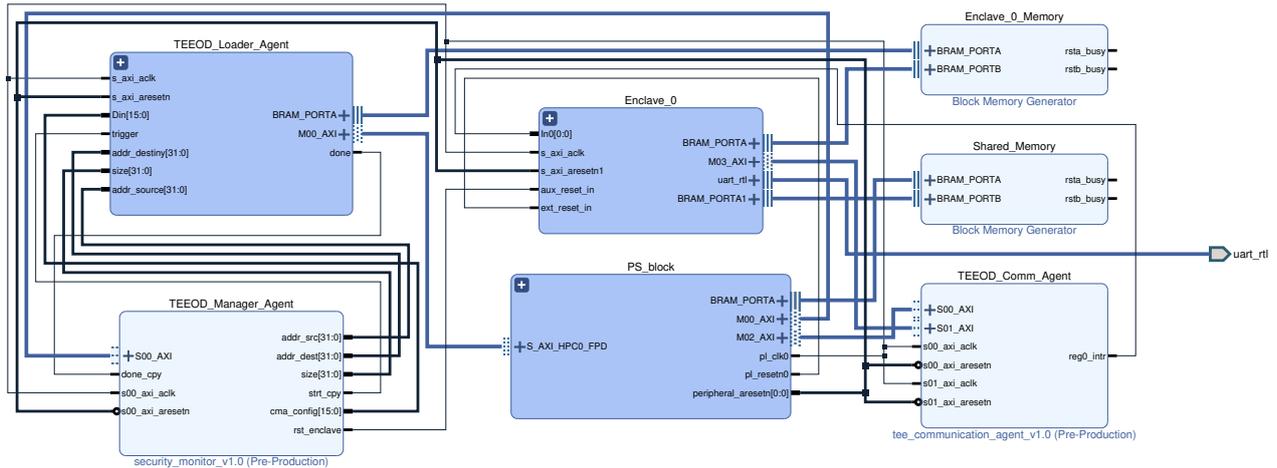

Fig. 5. TEEOD's block diagram with one enclave (Xilinx Vivado simplified view, i.e., clocks and resets hidden). "Enclave_0" is connected to a UART peripheral and two BRAMs. One to implement the shared memory mechanism and the other one to work as TCM, holding the TAs binary.

Next, we measure the execution time of the two different mechanisms present in the invoke-command operation (*TEEC_InvokeCommand*). In the first scenario, arguments are only passed through the message box. We test a simple TA that returns a value, provided by the CA, incremented by one. To evaluate communication via shared memory, in the second scenario we test a TA whose invoke-command outputs a 16-byte array to shared memory. We observe that the first scenario takes 201.0 microseconds on average to execute while the second one takes 327.05 microseconds on average.

Lastly, we tested the duration of the close session operation (*TEEC_CloseSession*). This operation took 40.1 microseconds to complete, which is the time of the REE to notify the TEEOD to close the session and receive a reply of whether it went right or wrong. All close session-related operations, i.e., operations to clear the enclave, are performed in a parallel way and obfuscated from the REE's point of view.

## VII. RELATED WORK

There are several security-oriented hardware architectures that have been leveraged to assist and facilitate the deployment of TEE systems.

*CPU-Extensions.* One common TEE design approach relies on dedicated CPU extensions, e.g., Intel SGX [33] and Arm TrustZone [6]. This approach adds hardware into the CPU to support isolation between execution environments, and the main advantage is lowering costs by reusing the CPU to execute in more than one security state. SGX has have been leveraged to implement confidential computing architectures in the cloud [34], [35], while TrustZone has found mainstream adoption in the mobile industry, providing a secure environment for third-party security sensitive operations [36]. However, due to the high number of disclosed vulnerabilities and attacks both in TrustZone [19], [36] and SGX [7], [8], the real security guarantees of these technologies have been put into question.

*On-Chip Secure Processors.* On-Chip secure processors have also been used to provide a secure execution environment. This way a secure processor is built into the SoC with dedicated memory, addressing the concerns of sharing hardware resources with the main CPU. Examples of such technology include Apple's SEP [37], Qualcomm's SPU [38], Intel ME [39], and HECTOR-V [18]. Typically, on-chip secure processors just provide a single TEE, limiting the broad applicability of the technology for several applications.

*External Secure Processors.* Google's Titan [16] and Apple's T2 [17] are external dedicated security processors mounted externally to the SoC. Other external security environments include Amazon's Graviton [40], which uses a GPU to isolate security sensitive operations from the main CPU; HETEE [41] which uses the PCIe connections to securely link untrusted CPUs to trusted isolated CPUs that interface with high-performance hardware peripherals; and TPM [42], commonly used security chips that store secretes and perform cryptographic operations used for attestation.

*FPGA-based TEEs* Recently, FPGAs start being used to create isolated environments for performing both software and hardware security sensitive operations. ShEF [21] and MeetGo [43] propose mechanisms to securely load custom hardware designs onto a remote FPGA. Additionally, the hardware executes securely in an environment where a malicious agent cannot interfere with the operation or steal secrets. Ambassy [44] creates secondary TEEs in the FPGA to execute hardware versions of TAs, converted through HLS. These new TEEs are managed by the primary TEE executing in the Arm TrustZone (secure world) and are intended for third-party developers to execute their applications. AMBASSY runs the software applications as hardware bitstreams, imposing to the user the need to translate software applications to hardware logic. This can hamper interoperability and legacy support of existing

TAs. In contrast, in TEEOD, a software TA can execute transparently, as long as it is compliant with standard Global Platform internal APIs.

VIII. CONCLUSION AND FUTURE WORK

In this work, we proposed TEEOD, a novel TEE design that leverages reconfigurable FPGA technology. TEEOD implements secure enclaves in the PL by instantiating a customized and dedicated security processor per application on a per-need basis. We implemented a PoC implementation on an Ultra96-V2 platform and conducted experiments that demonstrated TEEOD is able to host up to 6 simultaneous enclaves. To demonstrate the practicability of TEEOD in real-world applications, we successfully ran an open-source Bitcoin wallet. In the future work, we expect to implement local TA attestation, support the loading of encrypted TAs, improve scalability, and leverage dynamic partial reconfiguration (DPR) technology to instantiate and destroy enclaves.